\documentclass[twocolumn,aps,showpacs,floatfix]{revtex4}
\usepackage{graphicx,amsmath}
                                                                                                
\begin{document}
 \title{Spectral properties and isotope effect in strongly
 interacting systems:\\  Mott-Hubbard insulator and polaronic semiconductor}

\author{S. Fratini}
\affiliation{Laboratoire d'Etudes des Propri\'et\'es Electroniques des Solides,
CNRS - BP166 - 25, Avenue des Martyrs, F-38042 Grenoble Cedex 9}
\author{S. Ciuchi}
\affiliation{Istituto Nazionale di Fisica della Materia and
Dipartimento di Fisica\\
Universit\`a dell'Aquila,
via Vetoio, I-67010 Coppito-L'Aquila, Italy}
\date \today

\begin{abstract}
We study the electronic spectral properties in two examples of strongly interacting
systems:  a Mott-Hubbard insulator with additional electron-boson interactions, 
and a polaronic semiconductor. An approximate unified framework 
is developed for the high energy  part of the spectrum, in
which the electrons move in a random field determined by 
the interplay between magnetic and bosonic fluctuations.  
When the boson under consideration is a lattice vibration,   
the  resulting isotope effect on the spectral properties 
is similar in both cases, being strongly temperature and energy dependent, in
qualitative agreement with recent photoemission experiments in the cuprates. 
\end{abstract}
\maketitle

\section{Introduction}

A general feature of strongly interacting electron systems is a sizeable
suppression of the  metallic character, accompanied by a redistribution of the
spectral weight from the region close to the Fermi level towards higher
energies.  In the presence of a strong electron-electron repulsion, for
example,  incoherent excitations arise far from the Fermi energy in  the
so-called upper and lower Hubbard bands,  at an energy scale which is ruled by
the strength $U$ of the interaction.  A similar behavior is also
found in systems with strong  electron-lattice coupling, where the spectral
weight is transferred to a broad  peak located around the polaron binding energy
$E_P$. 

Recent photoemission experiments in the  high-Tc cuprates 
\cite{Lanzara01,KShenPRL04,Gunnarssoncondmat05,Zhou05}, which are generally
described in terms of purely electronic models due to the  proximity to a Mott
insulating phase,
have revealed the existence of an important electron-lattice coupling.  This has
given rise to intense theoretical work focusing on the  excitation spectra in
the presence of  electron-boson interactions, both in models 
with \cite{NagaosaPRL04,GunnarssonEPJB05,Fehske} and without
\cite{Verga,Hague,HohenadlerPRB03,DevereauxPRL04,Kaiji04,Korni-alex}
electronic correlations. 
Of particular interest are those works which focus on isotope
effects (IE), since they can disentangle the properties
which are directly related to the coupling to the lattice degrees of freedom.
For example, it has been found that 
an anomalous isotope effect on the effective mass
arises at the polaron crossover, 
signaling  the breakdown of the Migdal-Eliashberg adiabatic
approximation \cite{Paola}. 
An enhancement of the IE is also expected  in the proximity of a Mott
metal-insulator transition, i.e. at intermediate values of
the electron-electron repulsion \cite{StJ}. 
On the other hand,  a strong  electronic repulsion suppresses
the effects of the lattice dynamics  on the { low energy}
excitations near the Fermi level.
These persist at { high energy}, where they amount essentially
to a broadening of the electronic spectra \cite{GunnarssonEPJB05}.

In this work, we provide an approximate theoretical framework based on
the Coherent Potential Approximation (CPA), which qualitatively
describes the high-energy excitation spectra resulting from both
electron-electron and electron-boson interactions, treating the
fluctuations of magnetic and bosonic origin as a local {\it static} disorder.
Although simplified,
the present analytical treatment 
is able to account for the existence of broad incoherent  
peaks at high energy, which are a common characteristic of strongly interacting
electronic systems.

We analyze 
two extreme cases which can be important for our general
understanding of the problem: (i) a 
Mott-Hubbard insulator  at
half-filling, in the presence of an additional  local interaction of the
electrons  with dispersionless bosons and (ii) a polaronic
semiconductor,  where the physics is solely  determined by the electron-boson
coupling. In the former case, the validity of our approach relies 
on the separation of electronic and bosonic energy scales, 
which is achieved due to a strong electron-electron repulsion. 
In the latter case, the present CPA results are controlled by direct
comparison with an exact solution obtained by Dynamical Mean Field Theory (DMFT)
 \cite{depolarone}.

The consequences of the electron-boson coupling on the  dispersion and width of
the  high energy features in the spectral function  are calculated in both
situations,  in section II and III respectively. Particular attention is devoted
to the effect of a shift of the boson frequency,  as can be achieved through an
isotopic substitution  if the bosonic mode that couples to the electrons is a
lattice vibration, or more generally if it is a collective mode with a sizeable 
lattice component (in which case a non-trivial IE can still arise, provided
that  the frequency of the boson is modified by the  isotopic substitution). 
In the present approximation, the IE on the high energy part of the
spectrum turns out to be  qualitatively similar in the
Mott-Hubbard insulator  and in the polaronic semiconductor, being  mainly
determined by the strength of the electron-boson coupling (albeit slightly
enhanced by the presence of electronic correlations): in both
cases it is  strongly temperature and energy dependent,  in agreement  with the recently
measured IE in the high temperature superconductor
Bi$_2$Sr$_2$CaCu$_2$O$_{8+\delta}$ \cite{Gweon-Nat04,Gweonsub}.
A tentative
analysis of the experimental results, performed in section IV, is
compatible with the 
existence of a  moderate electron-phonon coupling  in the cuprates.

\section{Electron-boson coupling in a Mott-Hubbard insulator}

\subsection{Coherent Potential Approximation}
We study the following Holstein-Hubbard Hamiltonian:
\begin{eqnarray}
\label{eq:themodel}
H & = & \sum_{k,\sigma}\epsilon_k  c^{\dagger}_{k,\sigma}
c_{k,\sigma} + U \sum_{i} n_{i,\uparrow} n_{i,\downarrow} -
\nonumber \\
& & - g\sum_{i,\sigma} n_{i,\sigma}(a_i +a^{\dagger}_i) + \omega_0 \sum_i
a^{\dagger}_i a_i,\nonumber
\end{eqnarray}
where electrons in a band with dispersion $\epsilon_k $ mutually interact
through an on-site repulsion $U$, and are coupled locally to a dispersionless
bosonic mode of frequency $\omega_0$, with a strength $g$. We shall set the
energy units such that $\hbar=k_B=1$.
We calculate the spectral properties of the above model at half filling, in the
framework of the Coherent Potential Approximation (CPA)
\cite{HubbardIII,Vollhardt}. This is suitable for the Mott insulating phase at
large $U$,  but also gives a fair description of the high energy incoherent 
excitations in the correlated metal, provided that the upper and the lower Hubbard bands are
well separated from the low energy quasi-particle peak \cite{DMFTreview}. The
latter, however,  is not accessible within this theory, and the position of the
chemical potential remains undetermined except at half-filling.
Additional features that are not included in the present description are 
the coherent excited states that arise at low temperature,
(even in the insulating phase),
due to the quantum nature of the magnetic excitations, 
in an energy range $J\sim t^2/U$ around the Hubbard band edges 
\cite{Preuss,Dagotto,Nasu}, and whose dispersion is srongly
reminiscent of spin density waves. 

Bearing these limitations in mind, 
the success of the CPA is that, despite its formal simplicity,  it correctly
accounts for the high-energy scattering by the randomly distributed magnetic moments.  The
momentum-integrated Green's function has the form \cite{Vollhardt}: 
\begin{equation}\label{eq:CPAlambda=0}
  G(\omega)=\frac{1}{2} \left [\frac{1}{G_0^{-1}(\omega) -\frac{U}{2}}+
  \frac{1}{G_0^{-1}(\omega) +\frac{U}{2}} \right ]
\end{equation}
where $G_0$ is an effective propagator which takes
hopping processes into account.  It 
can be eliminated by introducing a local
self-energy $\Sigma(\omega)$ through the following self-consistency condition:
\begin{equation}\label{eq:selfcons}
  G(\omega)=\sum_k \frac{1}{\omega-\epsilon_k-\Sigma(\omega)}=
  \frac{1}{G_0^{-1}(\omega) -\Sigma(\omega)}
\end{equation}
(in the ``atomic'' limit, $G_0=1/\omega$ and the usual single
site propagator is recovered \cite{Vollhardt}). 
For each frequency $\omega$, 
we are left with a system of two
equations for the two complex unknowns $G(\omega)$ and $\Sigma(\omega)$.

In the presence of a local electron-boson interaction, equation
(\ref{eq:CPAlambda=0}) can be generalized by
introducing an additional  field $y$, which accounts for the random
distribution of electronic energies
due to the fluctuations of the bosons at different sites
\begin{equation}\label{eq:CPA}
  G(\omega)= \int  \frac{dy}{2} P(y) \left [\frac{1}{G_0^{-1}(\omega)
  -\frac{U}{2}-y}+
  \frac{1}{G_0^{-1}(\omega) +\frac{U}{2}-y} \right ].
\end{equation}

In a system where the on-site electron-electron repulsion directly  competes
with the attraction induced by the bosons, the formation of  bipolarons (and
the resulting strong anharmonicities in the boson field) 
\cite{Millis,polaronCO} is prevented provided that $U$ is much larger than the
polaron binding energy $E_P = g^2/\omega_0$ \cite{StJ,Hewson}.
In this case, we can assume that the boson  
field obeys the following gaussian distribution,
\begin{equation}
 \label{gaussian}
  P(y)=\frac{1}{\sqrt{2 \pi \sigma^2}} \exp \left[-\frac{y^2}{2\sigma^2}\right]
\end{equation}
The variance 
\begin{equation}
  \label{eq:variance}
  \sigma^2=E_P  \omega_0   \coth\left(\frac{ \omega_0}{2 T}\right)
\end{equation}
is determined either by the thermal fluctuations
(at high temperature, $T\gg \omega_0$) or by the quantum  fluctuations
of the bosons (at low temperature, $T\ll \omega_0$).

Let us emphasize that we have neglected boson renormalization effects which, 
even in the presence of a strong electron-boson interaction, 
are suppressed by a sufficiently strong repulsion $U$ \cite{StJ,Hewson}. 
Such effects can in principle be included by taking $\omega_0$ in
eq. (\ref{eq:variance}) equal to the measured boson frequency rather than the 
bare parameter of eq.  (\ref{eq:themodel}).
However, the present scheme is expected to break down in
situations where  the bosonic and electronic energy scales become comparable
(i.e. $E_P \simeq U$),
in which case the properties of the bosonic field depend crucially on
the electron correlations.

Equations  
(\ref{eq:selfcons}) and (\ref{eq:CPA}) form the backbone of our theory.
The latter can be further explicited by evaluating the Hilbert-transform of
the gaussian 
\begin{equation} \label{Hilbtrasf}
\int d y P(y) \frac{1}{z-y} = -i \sqrt{\frac{\pi}{2\sigma^2}} \
 {\mathcal W} \left[\frac{z}{\sqrt{2\sigma^2}}\right]
\end{equation}
where ${\mathcal W}$ is the complex error function \cite{abramowitz}.
Note also that, since the self-energy $\Sigma$ does not depend on
momentum,  the details of the band dispersion $\epsilon_k$ enter 
in equation (\ref{eq:selfcons}) only through the 
corresponding density of states. 
For the sake of simplicity, we shall consider a band dispersion
$\epsilon_k$ with a semicircular density of states of half-width $D$,
for which the self-consistency equation  (\ref{eq:selfcons}) reduces
to $G^{-1}_0=\omega-D^2 G/4$. However, since they rely on
a momentum independent  quantity (the local self-energy), 
the results 
are quite independent on the choice of the band dispersion.

The spectral function, which is the quantity of interest in the
present work, is defined as
\begin{equation}
A(\epsilon_k,\omega) = -\frac{1}{\pi} Im
\frac{1}{\omega+i\delta -\epsilon_k-\Sigma (\omega)}.
\end{equation}
Its momentum integral --- the spectral density --- 
is $N^*(\omega) =-\frac{1}{\pi} Im G(\omega)$.

\subsection{Mott-Hubbard insulator}

\begin{figure}[htbp]
  \centering
  \includegraphics[scale=0.4,angle=0]{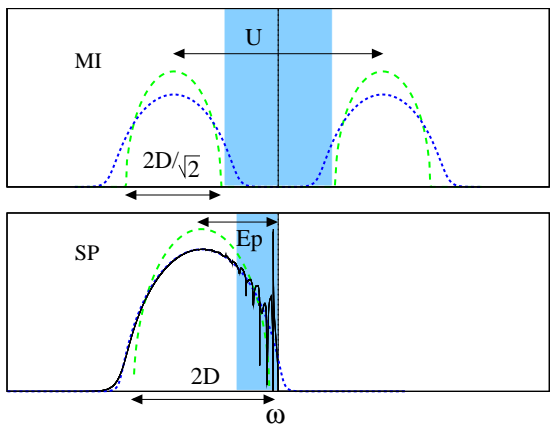}
  \caption{(color online)  A sketch of the spectral density (the
    momentum-integrated spectral function). Top: in a Mott-Hubbard
    insulator, for $U/D=3$  without (dashed green) and
  with (dotted blue)  electron-boson interaction ($E_P/D=0.9,
  \omega_0/D=0.1$). Bottom: in a polaronic system ($U=0$) with the same
  electron-boson parameters. The green dashed curve here is 
  the noninteracting band, the black solid line is the 
  exact DMFT result \cite{depolarone}, and the  blue dotted curve is the
  approximate result based on equation (\ref{eq:CPA-SP}). 
  Vertical lines mark the reference energy of
  incoming particles. In both panels the shaded area represents the low energy
  part of the spectrum.}
  \label{fig:DOS}
\end{figure}

For large $U$, and in the absence of electron-boson interactions, the
set of equations (\ref{eq:CPAlambda=0}) and (\ref{eq:selfcons})
can be rewritten as
\begin{eqnarray}
  \Sigma&=&\frac{U^2}{4 [\omega-\frac{D^2}{4}G]} \label{eq:solCPAel1}\\
  G&\simeq& \frac{1}{2}\frac{1}{\omega-\frac{D^2}{4}G \pm U/2}  
 \label{eq:solCPAel2}
\end{eqnarray}
 The spectral density 
$N^*(\omega)\simeq \frac{2}{\pi {D}^2} \sqrt{{D^2}/{2}-\left (
   \omega\pm U/2 \right)^2}$
is illustrated in figure 1.a, and
consists of two bands of reduced width $\simeq
2D/\sqrt{2}$ separated by a gap $\Delta \simeq U/2-D/\sqrt{2}$ 
(the $+$ and $-$ sign are for the lower and
upper Hubbard band  respectively).
Such bands represent incoherent states that are strongly scattered by the
disordered magnetic moments. The corresponding spectral function $A(\epsilon_k,\omega)$ 
exhibits broad peaks, whose
width is comparable with the bandwidth itself: it is determined 
by the scattering rate 
$\Gamma(\omega)=-Im \Sigma (\omega) \sim D^2
N^*(\omega)$, which  is proportional to the spectral density, and
is therefore  strongly energy dependent. 
As a consequence, the peaks sharpen (and become asymmetric)
when approaching the band edges, as can be seen in the
energy scans at constant momentum, in figure 2.a (dashed green curve).
Their  dispersion, 
defined as the locus of the maxima of $A(\epsilon_k,\omega)$ at constant
$\omega$ (the so-called momentum distribution curves, MDC) can be
obtained by solving the equation $E_k-Re \Sigma (E_k)=\epsilon_k$, and
is roughly given by $E_k=-U/2+\epsilon_k/\sqrt{2}$ (see fig. 2.b).

It should be noted that the slope  of
the dispersion of such high energy features deviates significantly
from the
unrenormalized Fermi velocity $v_F$. For this reason, 
in strongly correlated electron systems,
some care should
be taken when extracting the function $Re \Sigma(\omega)$ from experimental 
data  under the 
assumption that $v_F$ tends asymptotically to its
unrenormalized value at sufficiently high
energy, as is customary in the field \cite{Valla99} (see also
ref. \cite{Fink}).

\subsection{Electron-boson coupling}

The coupling to a bosonic mode leads to several modifications of the
picture described above. First of all, the fluctuations of the site
energies
lead, through the relation
(\ref{eq:CPA}), to an overall broadening of
the Hubbard bands. In particular, this  generates additional
exponential tails  in the vicinity of the band edges,
whose extension is governed by the variance $\sigma\sim \sqrt{E_P \omega_0}$.
In the adiabatic regime, where the boson frequency is small compared
to the bandwidth ($\omega_0\ll D$), and 
for large $U$, such tails are typically smaller than both the width of
the Hubbard bands and the size of the gap (see figure 1.a)

The spectral function $A(\epsilon_k,\omega)$,  for
$U/D=3$ and $\sigma=0.3 D$, is
illustrated in figure 2.a at different values of $\epsilon_k$. 
This value of the variance corresponds, for example,  to a moderate
electron-boson coupling $E_P/D=0.9$, 
in the adiabatic regime $\omega_0=0.1 D$ and at zero temperature.
The purely electronic case ($\sigma=0$) is also shown
for comparison (dashed green lines).
The electron-boson coupling
strongly alters the lineshapes:   
well inside the Hubbard bands, there is a huge broadening of the
peaks, which can be estimated  in the large $U$ limit (see appendix) to 
$\Delta \Gamma/\Gamma \sim 12 \sigma^2/D^2$, and is of the order of
$100$\%  in the present example. 
However, the effect is even more dramatic in
in the vicinity of the band edges. There,  the scattering rate, which
is roughly proportional to the spectral density, has a sharp 
(square root) dependence on the energy  in the pure electronic case, 
causing a marked asymmetry of the peaks.  
The  boson fluctuations convert the sharp edge in a 
much smoother exponential tail, restoring a more symmetric lineshape
for $A(\epsilon_k,\omega)$.
Let us mention that the above-mentioned power law behavior at the
edges of the Hubbard bands is not peculiar to
the semi-circular DOS used in this example  (see
ref. \cite{Brinkman}). Indeed, analogous results are obtained 
starting with  a flat noninteracting DOS with a step-like edge,
appropriate for two-dimensional systems.

\begin{figure}[h!tbp]
  \centering
\includegraphics[scale=0.38]{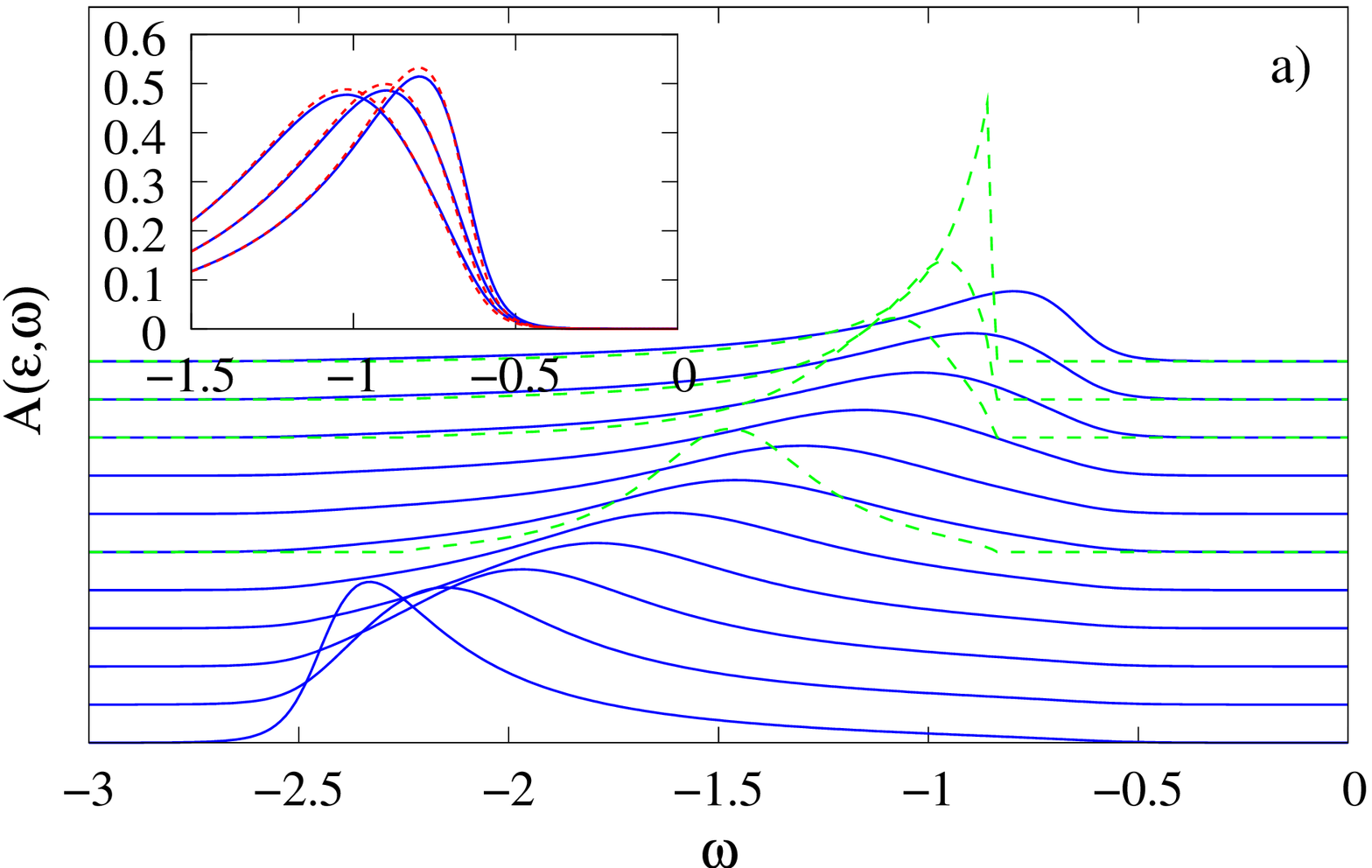}\\
\vspace{.5cm}
\includegraphics[scale=0.4]{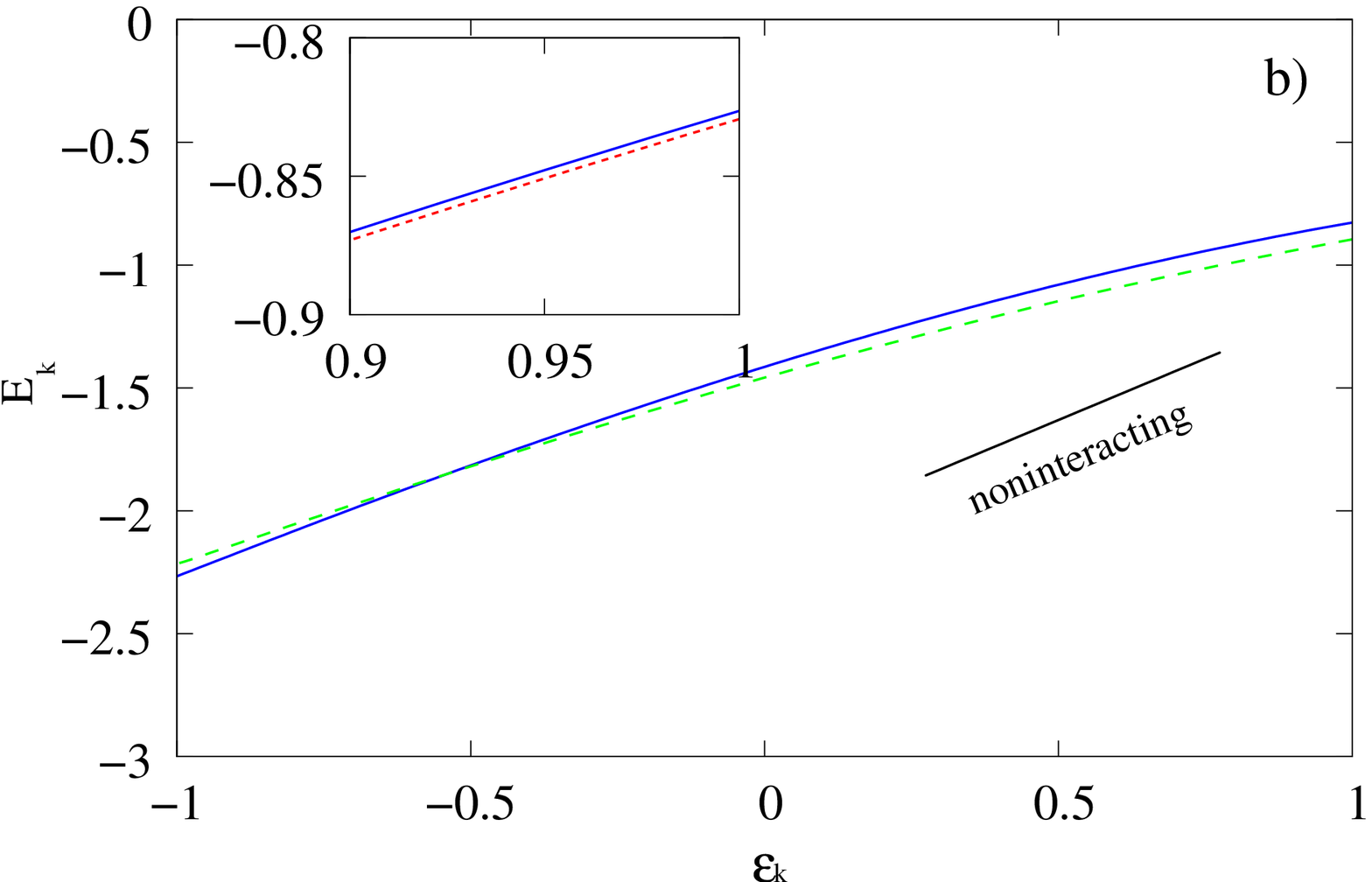}\\
\vspace{.5cm}
\includegraphics[scale=0.4]{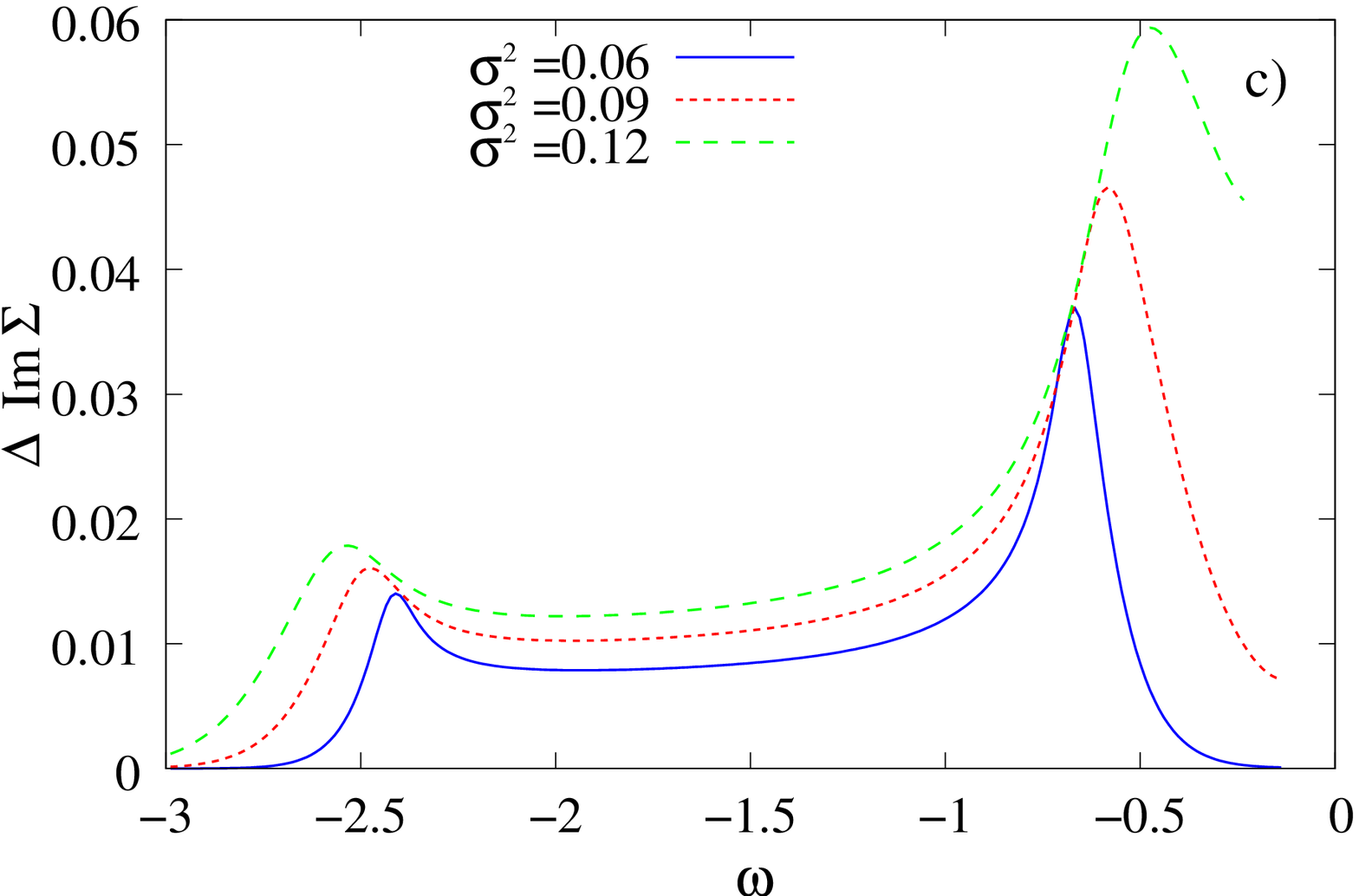}
\caption{(color online) From top to bottom: a)  energy scans of the spectral function
    $A(\epsilon_k,\omega)$ in a Mott-Hubbard insulator with additional
    electron-boson interactions, at
    different $\epsilon_k$ (equally spaced between the two edges
    $\epsilon_k=\pm D$) (solid blue lines). The parameters are $U/D=3$,
    $E_P/D=0.9$,$\omega_0/D=0.1$, and energies are in units of $D$. 
    The curves for $\epsilon_k/D=1,0.8,0.6,0.0$ in the absence of
    electron-boson coupling are shown for comparison (dashed green
    lines). The inset shows
    the effect of a shift $\Delta \omega_0/\omega_0=-6\%$ of the boson
    frequency on the spectral function at the same values of
    $\epsilon_k$ (the dotted red line is with the modified frequency).
    b) the  dispersion of the broad peaks deduced from momentum scans
    at constant energy, for $U/D=3$ with (solid blue line) 
    and without (dashed green line) electron-boson
    coupling. The slope of the noninteracting band is indicated.
    The inset shows the IE on the dispersion 
   (the dotted red line is with the modified frequency).
   c) Absolute value of the isotope effect on the scattering rate
   $\Gamma=-Im \Sigma$,   
   defined as $\Delta \Gamma=\Gamma[\omega_0+\Delta
   \omega_0]-\Gamma[\omega_0]$, for different values of
   $\sigma^2/D^2=0.06,0.09,0.12$.}
  \label{fig:MI}
\end{figure}

The position of the peaks is also affected by the electron-boson
coupling, although to a weaker degree (this is partly 
in contrast with the results of refs. 
\cite{NagaosaPRL04,Gunnarssoncondmat05}, which only
predict a broadening of the peaks, but no renormalization of the dispersion).
In figure 2.b, we have reported the renormalized 
MDC dispersion $E_k$.
The main effect of the electron-boson coupling is an increase of the
slope $v_{he}=d E_k/d\epsilon_k$; its relative variation is of the order
$ \sigma^2/D^2$, 
as can be estimated from the large $U$
expansion presented  in the appendix, and
is one order of magnitude smaller than  the corresponding effect on
the linewidths.

\subsection{Isotope effect on the spectral properties}

We shall now analyze the consequences of a change in the
boson frequency on the spectral properties.
Unless otherwise specified, 
we shall consider  a relative shift $\Delta\omega_0/\omega_0=-6$\%,
that can be  achieved through the substitution 
${}^{16}O\to {}^{18}O$ in the case of a lattice mode with
predominantly oxygen character \cite{Kovaleva}. The value inferred from
the photoemission experiments in ref. \cite{Gweon-Nat04} on the mode
at $\omega_0=70meV$  is comparable with this value,  $| \Delta\omega_0|/\omega_0
\gtrsim 7 $\% , pointing to a strongly phononic character of the bosonic
excitation.

By direct inspection of equation
(\ref{eq:variance}), we immediately see that any isotopic effect 
will be strongly
temperature dependent, on the scale of the boson frequency itself:
at low temperatures, where $\sigma^2= E_P \omega_0$,
a modification of $\omega_0$ directly affects the distribution (\ref{gaussian}) of 
the lattice displacements, and therefore
modifies the spectral properties described above. However, this effect
is rapidly suppressed 
when the thermal fluctuations
become dominant, in which case  
$\sigma^2= 2E_P T$ is independent on the boson frequency.
In the following, we shall present the results at $T=0$, where the IE
is maximum.  The IE at any temperature can be obtained
straightforwardly by multiplying the results by an appropriate
coefficient 
\begin{equation}
  \alpha(T)=\frac{\partial  \sigma^2}{\partial  \omega_0}=\left[
1-\frac{\frac{\omega_0}{T}}{\sinh \frac{\omega_0}{T}}\right]
\coth[ \frac{\omega_0}{2T}],
\end{equation}
which is maximum at $T=0$ ($\alpha=1$), and rapidly drops at temperatures
$T\gtrsim 0.2 \omega_0$, where $\alpha\sim \omega_0/3T$.

The inset of figure 2.a shows the IE on the peaks in
$A(\epsilon_k,\omega)$ at various $\epsilon_k$ close to the edge of the
LHB. The reduction of the boson frequency leads to a weak 
shift of the peak position, corresponding to a slight 
reduction of the slope $v_{he}$, which is too weak to be observed at 
such moderate values of $E_P$ 
(see also the inset of figure 2.b, where we have reported the
renormalized dispersion close to the upper edge of the band).
On the other hand, we know from
the previous section that  the effect of the electron-boson coupling 
is much stronger on the linewidths than on the dispersion, and indeed
some reduction of the peak widths is already visible in the plots of 
$A(\epsilon_k,\omega)$.
For a more quantitative analysis, we have reported in figure 2.c 
the IE on the scattering rate, defined as 
$\Delta \Gamma=\Gamma[\omega_0+\Delta\omega_0]-\Gamma[\omega_0]$. It 
shows an interesting dependence on the energy: 
it is rather flat inside the Hubbard bands, but exhibits
pronounced peaks of width $\sim \sigma$ around the band edges, because
this is the region where the spectra are mostly affected by the
bosonic fluctuations. 
The  variation of the linewidth in the flat region 
well inside  the Hubbard bands is 
approximately given by (see appendix)
\begin{equation}\label{eq:DGammamin}
  \frac{\Delta \Gamma_{min}}{\Delta \omega_0}\simeq 3 \sqrt{2} \frac{E_P}{D}
\end{equation}
and gives a direct measure of the strength of the electron-boson
coupling. Note that although the precise  
numerical coefficient is specific to the semi-circular
DOS considered here, analogous formulas can in principle 
be derived for any choice of the noninteracting band
(in fact, a similar result also holds in the weak coupling regime
\cite{Engelsb}).

Above the band edges, the IE decays exponentially, following the
fluctuation induced tails in the spectral density. This gives rise to
an extremely asymmetric peak of $\Delta \Gamma$ at the band edge, which becomes 
more and more symmetric as $\sigma$ is increased
(cf. the discussion on the
spectral function in the previous section).
For sufficiently large  $\sigma\gtrsim D$, 
$\Delta \Gamma(\omega)$ tends to a skewed gaussian, 
whose maximum is located in $\omega\simeq -U/2+2 \sigma$, whose width
is proportional to $\sigma$ and height scales as
\begin{equation}
  \frac{\Delta \Gamma_{max}}{\Delta \omega_0} \sim \sqrt{\frac{E_P}{\omega_0}}.
\end{equation}

\section{Polaronic semiconductor}

In this section, we use the approximate theory presented above 
to address the spectral properties in a system with electron-boson
interactions, but in the absence of electron-electron correlations. 
For this problem, extensive results are available in 
the litterature for the weak coupling regime \cite{Engelsb},
and for the polaronic anti-adiabatic regime, where the boson frequency
is assumed to be much larger than the noninteracting
bandwidth \cite{AlexRann}.  We shall focus instead on the
polaronic adiabatic regime (i.e. moderate to strong electron-boson
couplings $E_P\gtrsim D$ and  adiabatic bosons $\omega_0\ll D$), 
which is more often encountered in solids \cite{Dessau,Perfetti,Okazaki}, 
and for which a simple formulation of the spectral
properties is not clearly established.

For the problem of a single electron coupled to a
dispersionless boson, a complete characterization of the excitation
spectrum has been given in references \cite{depolarone,BRopt}, 
based on the Dynamical Mean Field Theory. 
In the  adiabatic regime,
the spectra are composed of one (or several, depending on the coupling
strength) narrow features at low energy, equally spaced by $\omega_0$, 
coexisting with a continuous high energy background centered around 
the polaron binding energy $E_P$.
As the coupling strength increases, 
the low-energy features are rapidly suppressed and 
the spectral weight becomes dominated by the high-energy incoherent
background.
It should be stressed that, contrary to what happens in the
anti-adiabatic limit, the high-energy features here are \textit{dispersive} due
to the strong hybridization with the free-electron states (see e.g. fig. 14
in ref. \cite{depolarone}). 

The high-energy incoherent excitations are well described by
an equation analogous to (\ref{eq:CPA}),
\begin{equation}\label{eq:CPA-SP}
  G(\Omega)= \int dy P(y) \frac{1}{G_0^{-1}(\Omega)-y},
\end{equation}
where the boson field obeys the same gaussian distribution of eq. (\ref{gaussian}).
In fact, the above equation can be shown to be \textit{rigorously}
valid in the framework of the DMFT
in the adiabatic limit $\omega_0=0$ [see ref. \cite{depolarone}, eq. (46)]. 
A similar relation also holds  for a system of
spinless polarons at finite density \cite{Millis,polaronCO}, leading
to very similar results as in the single particle case presented here
\footnote{In this case,  
  the expression (\ref{eq:CPA}) can be rigorously derived from the
  Holstein model in the classical limit $M\to \infty, T\gg \omega_0$.
  However, contrary to the single electron case, the distribution
  $P(y)$ is modified by the interaction with a \textit{finite density} of
  electrons. It becomes
  bimodal, with two peaks separated by  $\Delta \sim 2 E_P$,
  leading to an  upper and a lower ``Holstein band'' in the
  spectral density (in analogy with the upper and lower Hubbard bands
  in correlated systems).}.

Note that the chemical potential is undefined  in the present single
particle problem.
To make direct contact with the results of the previous section, we
shall interpret the results for the spectral function as corresponding
to the lower ``Holstein band'' (i.e. to excitations 
at negative binding energies as in a direct photoemission experiment),
setting the chemical potential at the extremum of the polaron band, with
the replacements $\omega=-\Omega+E_0$ and $\epsilon_k\to -\epsilon_k$ 
in equation (\ref{eq:CPA-SP}). Here
$E_0\le -D$ is the polaron binding energy taken from the
DMFT solution, which tends to $-E_P$ in the strong coupling limit
$E_P/D\gg 1$ (see fig. 4 in ref. \cite{depolarone}).

We shall first solve the coupled equations (\ref{eq:CPA-SP}) and
(\ref{eq:selfcons}) in a regime where the variance $\sigma$ 
of the boson field is smaller than the noninteracting
bandwidth, which is typically the case for moderate values 
of the electron-boson coupling. The opposite limit $\sigma\gtrsim D$,  where the 
spectral density takes the form of a gaussian, multi-boson shakeoff peak,
will  be treated  at the end of this section.

\subsection{Intermediate coupling regime}

\begin{figure}[htbp]
  \centering
\includegraphics[scale=0.38]{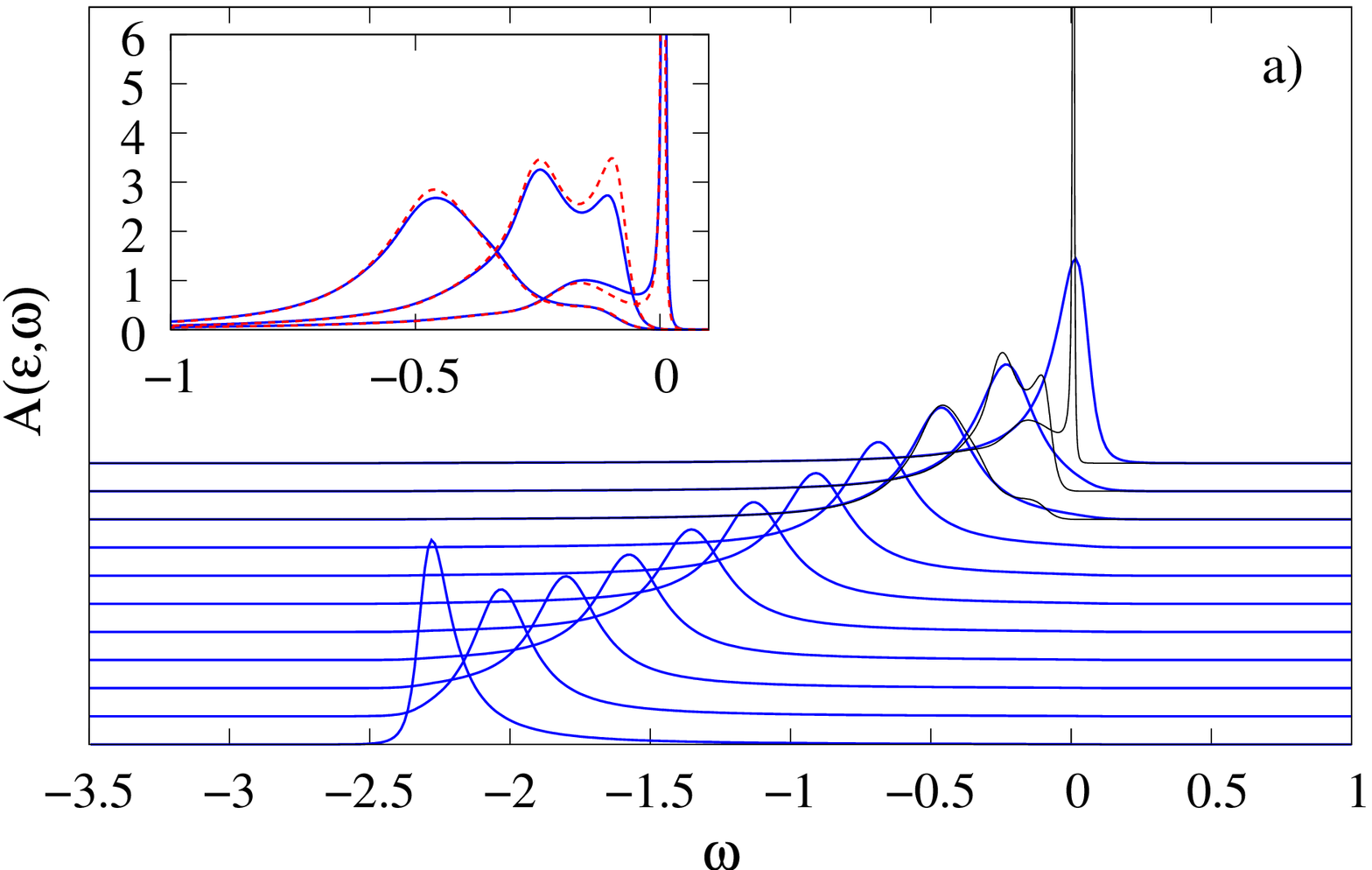}\\
\vspace{.5cm}
\includegraphics[scale=0.4]{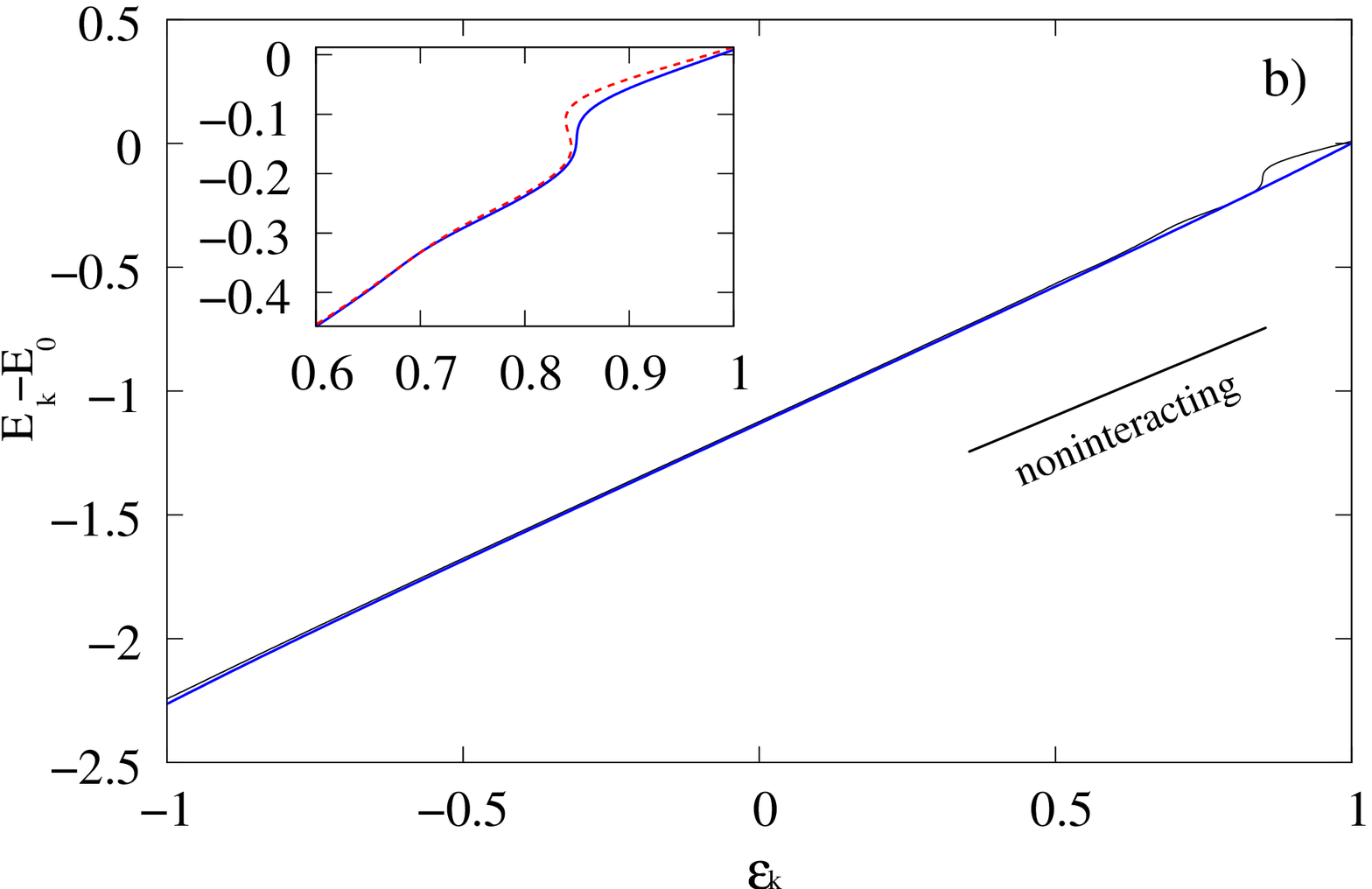}\\
\vspace{.5cm}
\includegraphics[scale=0.4]{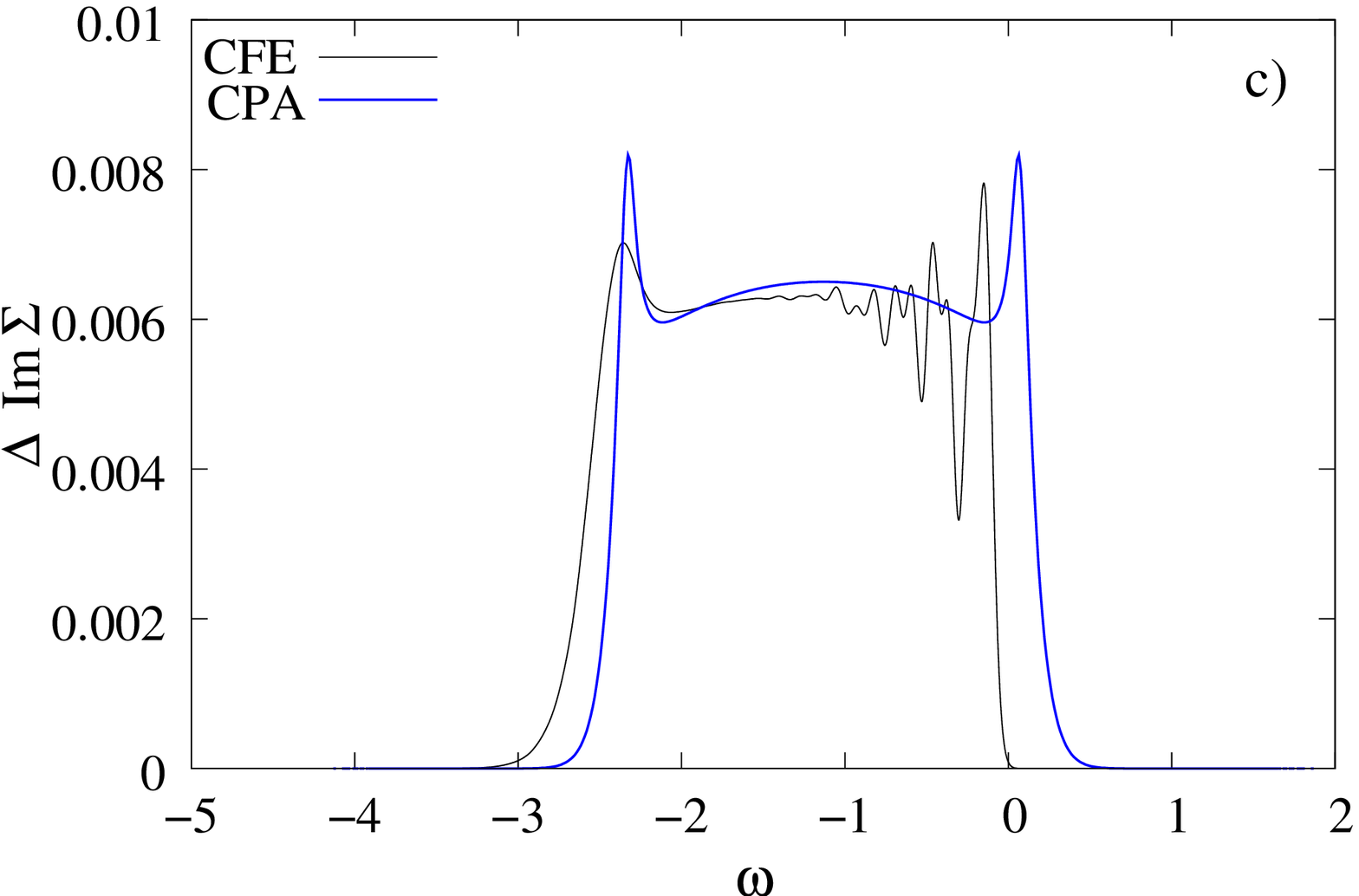}
  \caption{(color online) From top to bottom:  
  a) Energy scans of the spectral function
    $A(\epsilon_k,\omega)$ in a polaronic semiconductor, at
    different $\epsilon_k$ (equally spaced between the two edges
    $\epsilon_k=\pm D$) within approximation eq. (\ref{eq:CPA-SP}) (solid blue
    lines). 
    The parameters are 
    $E_P/D=0.9$,$\omega_0/D=0.1$. The curves for $\epsilon_k/D=1,0.8,0.6$, 
    obtained using the exact DMFT solution, 
    are shown for comparison (thin black lines). The inset shows
    the effect of a shift $\Delta \omega_0/\omega_0=-6\%$ of the boson
    frequency on the spectral function at the same values of
    $\epsilon_k$ (dotted red line is with the modified frequency).
    b) the  dispersion of the broad peaks deduced from momentum scans
    at constant energy using  DMFT (thin black line) 
    and the approximate theory (solid blue line). 
    The  slope of the noninteracting  band is indicated.
    The inset shows the IE on the dispersion obtained in DMFT 
   (the red dotted line is with the modified frequency).
   c) Absolute value of the isotope effect on the scattering rate
   $\Delta \Gamma$ using DMFT (thin black line) and the approximate theory 
   (solid blue line).
   Data from exact DMFT are shown after convolution with a gaussian 
   filter of width $0.05 D$.}
  \label{fig:SP}
\end{figure}

The spectral density for $E_P/D=0.9$ and $\omega_0/D=0.1$
($\sigma=0.3D$) is illustrated in figure 1.b (blue dotted line).
As in the correlated case, 
the boson fluctuations result in an overall broadening of the original
band (green dashed line).  
The  agreement with the  DMFT result (black solid line) is excellent in the smooth
region at high binding energies, while in the low energy region, 
the CPA clearly misses the detailed structure of the narrow
peaks. However, even there it 
gives a fair description of the integrated spectral weight, 
which is what one would measure experimentally in the presence of 
a sufficient  energy broadening.

\subsubsection{High-energy features}
The results for the spectral function
are illustrated in figure 3.a.
As in the correlated case treated in the previous section, the high
energy part of the spectra
is characterized by broad features, whose dispersion  (fig. 3.b)
sensibly deviates from the noninteracting case.
The variation of the scattering rate
under a shift of the boson frequency (fig. 3.c) is also
 very similar to the correlated case, corroborating the fact that 
the magnetic  fluctuations only play an indirect role
in the isotope effect: $\Delta \Gamma(\omega)$ is
rather flat at the center of the band, and attains its maximum 
in a narrow region of width $\sigma$ around the band edges.

For a more quantitative understanding,
analytical expressions 
can be obtained by an expansion to lowest order in the variance, as was
done previously for the Mott-Hubbard insulator. In the present case the
coupled equations   (\ref{eq:CPA-SP}) and
(\ref{eq:selfcons}) become
\begin{eqnarray}
  \Sigma&\simeq &\sigma^2 G \\
  G&\simeq &\frac{1}{\Omega-(D^2/4+\sigma^2)G}
\end{eqnarray}
which is valid inside the band, far from the edges.
The Green's function has the noninteracting form, but with a
renormalized bandwidth given by $D\to \sqrt{D^2+4\sigma^2}$.
The slope of the dispersion of the high-energy features is renormalized
accordingly:  
$v_{he}=dE_k/d\epsilon_k=1/(1-2 \sigma^2/D^2)$, i.e. it 
does not coincide with the value in the absence of interactions, as
can be seen in figure 3.b.  The boson induced variation  is of the same
order as what was calculated in section II.C
in the presence of electronic correlations.

From the first equation we see that the scattering rate within the
band is proportional to the spectral density,
$ \Gamma(\Omega)\simeq \pi \sigma^2 N^*(\Omega)$.
Note that it is  weaker  than in the correlated case,
due to the absence of magnetic disorder.
Nevertheless, the modification of the scattering rate under a shift of the
boson frequency is of the same order as in the previous case
[although with a smaller prefactor, cf. eq. (\ref{eq:DGammamin})], namely
\begin{equation}
  \frac{\Delta\Gamma}{\Delta \omega_0} =  \frac{2E_P}{D},
\end{equation}
being directly proportional to the strength of the electron-boson
coupling.

The existence of a maximum of $\Delta\Gamma$ near the band edge
(figure 3.c) also compares well with the DMFT result,
although its actual position 
is slightly shifted to higher binding energies.
This can be understood by observing that 
the true edge of the incoherent dispersion is at $\omega=-\omega_0$
(not at $\omega=0$), which marks the boundary between the high-energy
and low-energy regions in the excitation spectra (see below).
Note that $\Delta \Gamma$ rapidly drops in the region $-\omega_0<\omega<0$,
where $\Gamma$ itself is extremely small due to our assumption of
dispersionless (gapped) bosons.

\subsubsection{Low-energy features}

The low-energy part of the spectral function is shown in the inset of
figure 3.a. The DMFT result shows that the broad incoherent peak progressively
disappears when the band edge is approached (i.e. at low momentum
transfers), while a narrow ``quasi-particle'' peak arises at binding
energies $|\omega|<\omega_0$. The
evolution of such peak-hump structure, which is characteristic of the
intermediate coupling regime (in the strong
coupling regime,  the narrow features are too weak to be observed) 
causes a discontinuous jump, or kink,  in the
dispersion, which clearly separates the high and low-energy regions
with different slopes,  
as illustrated in figure 3.b (see also figure 3 in
reference \cite{BRopt}). 
The isotope effect on the
kink region is shown in the inset. Note that in this plot, the isotope shift
vanishes at $\omega=0$ by definition, since the origin of energies has
been shifted to coincide with the band edge (see the discussion at
the beginning of this section).

\subsection{Strong coupling limit}

At extremely large values of the coupling strength $E_P/D\gtrsim
D/\omega_0\gg 1$ (or at sufficiently high temperatures $T/D\gtrsim D/E_P$), 
the variance of the boson  field can become comparable with the noninteracting
bandwidth ($\sigma\gtrsim D$). The latter  can  therefore be
neglected in eq. (\ref{eq:CPA}), replacing $G_0^{-1}=\Omega$.
Note that this 
does not correspond to the usual atomic, anti-adiabatic limit \cite{AlexRann},
where it is assumed from the beginning
that $D\to 0$ is the smallest energy scale in the
problem, 
resulting in \textit{dispersionless} high energy features.  
The present theory is valid in the opposite limit, $D\gg \omega_0$,
which is more often realized in solids \cite{Dessau,Perfetti,Okazaki}.
Due to the large transfer integrals between
molecular units, the discrete shakeoff spectrum
characteristic of isolated molecules is converted here into a continuous
gaussian spectral density \cite{depolarone},
and a sizeable high-energy dispersion is recovered.

We recognize from eq. (\ref{eq:CPA-SP})  that the spectral
density in this case coincides with the gaussian distribution 
 itself, 
 \begin{equation} \label{eq:spect-strong}
   N^*(\omega)=P(\omega+E_P)
 \end{equation}
whose width is governed by the variance $\sigma$ (we have performed
the shift $\Omega=-\omega-E_P$ using  the
fact that $E_0\to -E_P$ in the strong coupling limit).
The full Green's function $G$ can be read directly 
from eq. (\ref{Hilbtrasf}):
  \begin{equation} \label{eq:Gstrong}
G(\omega) = -i \sqrt{\frac{\pi}{2\sigma^2}} \
 {\mathcal W} \left[\frac{-\omega-E_P+i\delta}{\sqrt{2\sigma^2}}\right]
\end{equation}

At the center of the polaron peak, the self-energy
$\Sigma(\omega)=-\omega-E_P-1/G(\omega)$ 
tends to
\begin{equation}
  \Sigma(\omega)=-(\omega+E_P) \left(1-\frac{2}{\pi}\right) 
-i \sqrt{\frac{2 \sigma^2}{\pi}}
\end{equation}
As a consequence, 
the dispersion of the broad peaks in $A(\epsilon_k,\omega)$ 
tends to $E_k=-E_P+ (\pi/2)\ \epsilon_k$, 
which  defines an apparent bandwidth $=\pi D $ for the
incoherent features,  sensibly larger than the noninteracting
value $2D$ [this should not be confused with the width of the polaron
peak in the momentum-integrated 
spectral density of eq. (\ref{eq:spect-strong}), 
which is governed by the variance $\sigma$]. 
Correspondingly, 
the slope of the high-energy dispersion saturates to a finite
value in the strong electron-boson coupling limit, 
which is independent of the coupling
strength, and is larger than the noninteracting value
(cf. the situation in the Mott-Hubbard insulator at strong $U$, 
section II.B). 

In the energy interval spanned by the dispersion $E_k$, 
the scattering rate is roughly constant and directly proportional
to $\sigma$. Its variation under  
a change of the boson frequency 
is given by
\begin{equation}
  \frac{\Delta \Gamma}{\Delta \omega_0}=\sqrt{\frac{E_P}{2\pi \omega_0}}.
\end{equation}
which is similar to the behavior encountered near the edges of 
the Hubbard bands [cf. eq. (12)].

\section{Concluding remarks}

In this work, we have presented an approximate analytical  
theory which addresses the high
energy spectral properties  in systems characterized by a
strong  electron-boson coupling, 
both in the presence and in the absence of
electronic correlations. 
Concerning the electron-boson interaction alone, 
the present approach gives accurate results 
in the adiabatic regime (i.e. opposite to the standard strong-coupling
polaron theories), where its validity can be controlled by direct comparison with the
results of the Dynamical Mean Field Theory. For the correlation part, on
the other hand, it
reduces to the CPA treatment of reference \cite{HubbardIII}, which 
qualitatively accounts  for the incoherent high-energy excitations
located in the upper and lower Hubbard bands.

Although the microscopic mechanisms in the cuprates 
certainly go beyond the simple model and approximations presented here,
our results reproduce at least qualitatively several characteristics of the observed
photoemission spectra, such as the 
existence of broad peaks
with a sizeable momentum dispersion, which sharpen and become more
asymmetric (in the EDC scans) as the band edge is approached.

More specifically, we have calculated  
the isotope effect on the high-energy spectral features.
In the present framework, where the bosonic and magnetic fluctuations are
effectively decoupled, 
the overall behavior 
in the Mott insulator
and in the polaronic semiconductor is qualitatively similar.
The most notable result is the existence of 
a strongly energy dependent IE on the \textit{linewidths},
which is
maximum in a
narrow energy interval of width $\sigma\sim \sqrt{E_P \omega_0}$ 
near the band edges,
because this is the
region where the excitation spectrum is mostly affected by the electron-boson
interaction (new spectral weight is created there, due to the 
presence of boson fluctuations).

The present approach also predicts a strong temperature
dependence of the IE, which should be strongly suppressed
 when the temperature reaches  some fraction of the boson
frequency (typically $T\gtrsim 0.2 \omega_0$). 
This occurs because at high temperatures, the fluctuations
of the boson field are dominated by thermal effects, which are
independent on the boson frequency, and is in no way related to the
existence of a temperature dependent electron-boson coupling.
Note that, according to the above general arguments, 
an analogous energy and temperature dependence of the IE can also be
expected  in more accurate treatments of the electron correlations.

The results of section II can be tentatively compared with the experimental
results of references \cite{Gweon-Nat04,Gweonsub}.
If we associate the variance $\sigma$ 
with the observed width $\sim 0.2eV$ of the active IE
region, and take the value
$\omega_0=0.07eV$ for the boson energy, a ``polaron''
binding energy $E_P\sim 0.5 eV$ is obtained. Assuming a
noninteracting bandwidth  of the order of $1eV$ places this value in the
intermediate electron-boson coupling regime. 
The same value of $\sigma$ is also compatible with the magnitude of
the observed IE on the scattering rate. 
From the frequency softening $\Delta\omega_0= 5-10 meV$ deduced from
the shift of the kink energy, 
the theory predicts a decrease
of the scattering rate around the band edges of the order 
 $\Delta \Gamma \simeq  \sigma \Delta\omega_0/\omega_0\sim 10-30 meV$ (see the end of
section II.D), in agreement with  the experimental observations of
ref. \cite{Gweonsub}.

It should be stressed that the present theory, based on a momentum independent
electronic self-energy, clearly fails in addressing the strongly anisotropic
\textit{dispersion} observed in the cuprate superconductors, which demands to
go beyond a local (and classical) treatment of the bosonic and
magnetic fluctuations.  
In particular,  the edges of the Hubbard bands at low temperature 
are strongly affected by the ``spin density wave'' dispersion, 
especially at intermediate values of $U$. 
Another point which is beyond the range of validity of the present approach
is the enhancement of the low energy IE due to proximity to a Mott
metal-insulator transition. In this
case, even a weak electron-boson interaction may produce a huge effect
when the boson frequency is varied \cite{StJ}.

\acknowledgments
We are grateful to G.H. Gweon and A. Lanzara for sharing with us
their data prior to publication. We also thank E. Cappelluti
for useful discussions.

\appendix

\section*{APPENDIX: IE in the large $U$ limit.}

\paragraph{Inside the Hubbard bands.}
Well inside the LHB, we can  expand (\ref{Hilbtrasf}) for small
$\sigma^2$, and neglect the effect of the UHB in equation
(\ref{eq:CPA}) for sufficiently large $U$. In the case of a
semi-circular DOS, we obtain
\begin{eqnarray}
  G&\simeq&\frac{1/2}{\omega-D^2G/4+U/2} \left\lbrace 1+
    \frac{\sigma^2}{[\omega-D^2G/4+U/2]^2} \right\rbrace \nonumber  \\
&\simeq&  \frac{1/2}{\omega-(D^2/4+2 \sigma^2) G+U/2} \label{sameexp}
\end{eqnarray}
where
we have used the fact that
$\omega-D^2G/4+U/2=1/(2G)$   to lowest order in $\sigma^2$.
The solution for $G$ reads
\begin{equation}
    G(\omega)= \frac{1}{{D^*}^2}
  \left[\omega+U/2-\sqrt{\left ( 
   \omega+U/2 \right)^2-{D^*}^2}\right ]
\end{equation}
with ${D^*}^2=D^2/2+4\sigma^2$.
From the self-consistency relation $G^{-1}=\omega-D^2G/4 -\Sigma$ we obtain 
\begin{equation}
  \Sigma(\omega)\simeq -\omega-U+(D^2/4+4\sigma^2) \ G(\omega)
\end{equation}
The scattering rate
$\Gamma(\omega)=-Im \Sigma(\omega)$ is therefore directly proportional to the
spectral density at this energy. 
At the center of the band, it is given by 
\begin{equation}
  \Gamma= \frac{D \sqrt{2}}{4}\left[ 1+\frac{12 \sigma^2}{D^2}\right]
\end{equation}
Its isotope effect at $T=0$ 
\begin{equation}
  \frac{\Delta \Gamma}{\Delta \omega_0}=  3 \sqrt{2}\frac{E_P}{D}
\end{equation}
gives a direct measure of the strength of the electron-boson interaction.

\paragraph{Fluctuation induced tails.}

In the gap region between the Hubbard bands,  additional spectral weight is created
by the boson fluctuations. 
For small $\sigma$ and 
sufficiently far from the band edges, both terms in 
 (\ref{eq:CPA}) can be replaced by their atomic counterparts
leading to the following exponential decay (for the LHB)
\begin{equation}
  \Gamma(\omega)\simeq \sqrt{\frac{\pi}{8 \sigma^2}} 
  \frac{[\omega^2-U^2/4]^2}{\omega^2}
  e^{-\frac{\left[\omega+U/2-
\frac{D^2\omega/4}{\omega^2-U^2/4 }\right]^2} {2\sigma^2}}
\end{equation}
(note that this formula is not valid close to the band edge, i.e. where 
the IE is maximum).

A simpler result is obtained for sufficiently large $\sigma$, i.e. in  
the strong electron-boson coupling regime $U\gg
\sigma^2\gtrsim D$. In this case, the variation $\Delta
\Gamma(\omega)$ of the scattering rate
under a shift of the boson frequency
takes the form of a skewed gaussian, which is
maximum at $\omega\simeq -U/2+2\sigma$, where  
\begin{equation}
 \frac{ \Delta\Gamma_{max}}{\Delta \omega_0}\sim \sqrt{\frac{E_P}{\omega_0}}
\end{equation}


\begin{thebibliography}{99}

\bibitem{Lanzara01} A. Lanzara, et al., Nature 412, 510 (2001).
\bibitem{KShenPRL04} K. M. Shen, et al., Phys. Rev. Lett. 93, 267002 (2004).
\bibitem{Zhou05} X. J. Zhou et al., cond-mat/0405130
\bibitem{Gunnarssoncondmat05}  O. R\"osch, et al., cond-mat/0504660
\bibitem{GunnarssonEPJB05} O. R\"osch, and O. Gunnarsson,
  Phys. Rev. Lett. \textbf{93} 237001 (2004);   \textit{idem}, 
  Eur. Phys. Journal B 43, 11 (2005). 
\bibitem{NagaosaPRL04} A. S. Mishchenko and N. Nagaosa,
  Phys. Rev. Lett. 93, 
036402 (2004).
\bibitem{Fehske} H. Fehske et al., Phys. Rev. B 69, 165115 (2004)
\bibitem{Verga} S. Verga, A. Knigavko, and F. Marsiglio, 
Phys. Rev. B 67, 054503 (2003).
\bibitem{Hague} J. P. Hague, J. Phys. Condens. Matter 15, 2535 (2003)
\bibitem{HohenadlerPRB03}M. Hohenadler, M. Aichhorn, and W. von der
  Linden, Phys. Rev. B 68, 184304 (2003). 
\bibitem{DevereauxPRL04}T. Cuk, et al., Phys. Rev. Lett. 93, 117003 (2004).
\bibitem{Kaiji04}  K. Ji, H. Zheng, and K. Nasu, Phys. Rev. B 70, 085110 (2004).
\bibitem{Korni-alex} P. E. Kornilovitch and A. S. Alexandrov
  Phys. Rev. B 70, 224511 (2004)  
\bibitem{Paola} P. Paci, M. Capone, E. Cappelluti,
S. Ciuchi, C. Grimaldi and
L. Pietronero Phys. Rev. Lett. {\bf 94}, 036406 (2005).
\bibitem{StJ} G. Sangiovanni, M. Capone, C. Castellani, and M. Grilli
Phys. Rev. Lett. {\bf 94}, 026401 (2005).
\bibitem{depolarone} S. Ciuchi, F. de Pasquale, S. Fratini and D. Feinberg, 
Phys. Rev. B {\bf 56}, 4494 (1997).
\bibitem{Gweon-Nat04} G.-H. Gweon, et al., Nature 430, 187 (2004).
\bibitem{Gweonsub} G.-H. Gweon, et al., submitted
\bibitem{HubbardIII} J. Hubbard, Proc. R. Soc. Lond. A281, 401 (1964); 
R. J. Elliott, J. A. Krumhansl, and P. L. Leath
Rev. Mod. Phys. 46, 465 (1974)
\bibitem{Vollhardt} D.Vollhardt in {\it Correlated Electron Systems} ed. by
V.J. Emery  (World Scientific, Singapore, 1992)
\bibitem{DMFTreview} A. Georges, {\it et al.} Rev. Mod. Phys. {\bf 68} 13 (1996).
\bibitem{Preuss} 
N. Bulut, D. J. Scalapino and S. R. White, Phys. Rev. Lett. 73, 748 (1994);
R. Preuss, W. Hanke and W. von der Linden, Phys. Rev. Lett. 75, 1344 (1995)
\bibitem{Dagotto} M. Ulmke, R. T. Scalettar, A. Nazarenko and
  E. Dagotto, Phys. Rev. B 54, 16523 (1996)
\bibitem{Nasu} N. Tomita and K. Nasu, Phys. Rev. B 60, 8602 (1999)
\bibitem{Millis} A.J.Millis, R. Mueller and B. I. Shraiman, Phys. Rev. B {\bf 54} ,5389,
(1996).
\bibitem{polaronCO} S. Ciuchi and F. de Pasquale Phys. Rev. B {\bf 59}, 5431 (1999).
\bibitem{Hewson} W. Koller, D. Meyer, and A. C. Hewson
Phys. Rev. B {\bf 70}, 155103 (2004).
\bibitem{abramowitz} {\it Handbook}  {\it of Mathematical} 
{\it Functions}
ed. by M. Abramovitz and I. Stegun (Dover, New York, 1964).
\bibitem{Valla99}T. Valla, {\it et al.} Science 285,  2110 (1999);
P. D. Johnson, et al., Phys. Rev. Lett. 87, 177007 (2001).
\bibitem{Fink}  A.A. Kordyuk, S.V. Borisenko, A. Koitzsch, J. Fink,
  M. Knupfer, H. Berger, Phys. Rev. B 71, 214513 (2005)
\bibitem{Brinkman} W. F. Brinkman and T. M. Rice, Phys. Rev. B 2, 1324
  (1970); M. E. Fisher and W. J. Camp, Phys. Rev. B 5, 3730 (1972)
\bibitem{Kovaleva} N.N. Kovaleva, et al., Phys. Rev. B 69, 054511 (2004)
\bibitem{Engelsb} S. Engelsberg and J.R. Schrieffer, Phys. Rev. {\bf 131}, 993
(1963).
\bibitem{BRopt}S. Fratini, F. de Pasquale, and S. Ciuchi, 
Phys. Rev. B 63, 153101 (2001)
\bibitem{AlexRann}
A.S. Alexandrov and J. Ranninger, Phys. Rev. B \textbf{45}, 13109 (1992);
J. Ranninger Phys. Rev. B {\bf 48}, 13166 (1993); G. J. Kaye,
Phys. Rev. B 57, 8759 (1998)
\bibitem{Dessau} D. S. Dessau et al., Phys. Rev. Lett. \textbf{81}, 192 (1998);
V. Perebeinos and P.B. Allen, Phys. Rev. Lett. 85, 5178 (2000)
\bibitem{Perfetti} L. Perfetti et al., Phys. Rev. Lett. \textbf{87}, 216404 (2001);
 L. Perfetti et al., Phys. Rev. B \textbf{66}, 075107 (2002)
\bibitem{Okazaki} A. Fujimori et al., J. Phys. Chem. Solids 57, 1379 (1996);
 K. Okazaki et al., Phys. Rev. B \textbf{69}, 165104 (2004)
\bibitem{Schrupp} D. Schrupp et al., cond-mat/0405623, accepted for publication in 
Europhys. Lett


\end{thebibliography}
\end{document}